\documentclass[prd,aps,floats,epsfig,twocolumn]{revtex4}
\usepackage{graphicx}

\newcommand{\beq}{\begin{equation}}

\newcommand{\eeq}{\end{equation}}

\newcommand{\bea}{\begin{eqnarray}}

\newcommand{\eea}{\end{eqnarray}}

\newcommand{\PTP}{\it Prog. Theo. Phys.}

\newcommand{\ApJ}{{\it Astrophys. J.\,}}

\newcommand{\PR}{{\it Phys. Rev.\,}}

\newcommand{\PRL}{{\it Phys. Rev. Lett.\,}}

\newcommand{\PL}{{\it Phys. Lett.\,}}

\newcommand{\etal}{{\it et al.}}

\newcommand{\eg}{{\it e.g. }}
\newcommand{\ie}{{\it i.e. }}

\begin{document}
\date{\today}

\title{Probing dark energy perturbations:\\ the dark energy equation of state and  speed of sound as measured by WMAP}
\author{Rachel Bean, Olivier Dor\'{e}}
\address{Department of Astrophysical Sciences, Princeton University, Peyton Hall - Ivy Lane, Princeton, NJ08544-1001, USA\\
  rbean@astro.princeton.edu, olivier@astro.princeton.edu} 
\date{\today}
\begin{abstract}
We review the implications of having a non-trivial matter component in
the universe and the potential for detecting such a component through
the matter power spectrum and ISW effect. We adopt a phenomenological
approach and consider the mysterious dark energy to be a cosmic
fluid. It is thus fully characterized, up to linear order, by its
equation of state and its speed of sound. Whereas the equation of
state has been widely studied in the literature, less interest has
been devoted to the speed of sound. Its observational consequences
come predominantly from very large scale modes of dark matter
perturbations ($k < 0.01 h\mathrm{Mpc}^{-1}$). Since these modes have
hardly been probed so far by large scale galaxy surveys, we
investigate whether joint constraints that can be placed on those two
quantities using the recent CMB fluctuations measurements by WMAP as
well as the recently measured CMB large scale structure
cross-correlation. We find only a tentative 1$sigma$ detection of the speed of sound, from CMB alone, $c_{s}^{2}<0.04$ at this low significance level. Furthermore, the current uncertainties in bias in the matter power spectrum preclude any constraints being placed using the cross correlation of CMB with the NVSS radio survey. We believe however that improvements in bias through improved survey scales and depths in the near future will improve hopes of detecting the speed of sound.
\end{abstract}

\maketitle

\section{Introduction}\label{sec1}

With the recent unveiling of the Wilkinson Microwave Anisotropy Probe
(WMAP) results, measuring the Cosmic Microwave Background (CMB)
anisotropy \cite{Bennett03}, the on-going supernovae searches
\cite{super} and the upcoming completion of the Sloan Digital Sky
Survey, amongst others, we are seeing a wealth of precision
observational data being made available. To a great extent the
standard $\Lambda$-CDM scenario fits the data well \cite{Spergel03}.  

However, the WMAP data 
might suggest that some modifications to the standard scenario are needed. 
One possible hint at required modifications 
is the deficit of large scale power in the temperature map, and in
particular, the low CMB quadrupole whose posterior probability is less
than a few hundredth (see \eg for possible interpretations
\cite{Efstathiou03, Contaldi03} and \cite{Efstathiou03b} for a
discussion of this number). 
One possibility is that this lack of large scale power might point to
some particular properties of the dark energy. The dominant
contribution to fluctuations on these scales is the Integrated Sachs Wolfe (ISW) effect
which describes the fluctuations induced by the passage of CMB photons
through the time evolving gravitational potential associated to nearby
($z<5$) large scale structures (LSS).  
One property we expect of dark energy is that it suppresses the
gravitational collapse of matter at relatively recent times, which in turn
suppresses the gravitational potential felt by the photons, thereby
leaving a signature in the ISW correlations. Since this signature is
created by the time evolving potential associated with relatively
close LSS, it should be intimately correlated with any tracer of LSS
\cite{Kofman85,Crittenden96}.  A positive detection of such a cross
correlation using WMAP data, assuming a cosmological constant as the
dark energy, has recently been measured
\cite{Boughn03,Nolta03,Fosalba03}.  

However the underlying cause of the dark energy is still unknown; and
such observational inferences offer rich prospects for guiding and
leading the theoretical effort. A wide variety of models have been
proposed to explain observations, from the unperturbed cosmological
constant, to a multitude of scalar field ÒquintessenceÓ and exotic
particle theories (see \cite{Peebles02} for a review). 

Much effort has been put into determining the equation of state of
dark energy, in an attempt to constrain and direct theories. Since the
equation of state affects both the background expansion and the evolution of
matter perturbations there are a wealth of complementary observations
available (again see \cite{Peebles02} and references therein). 

An equally insightful, but less investigated, characteristic of dark
energy is the speed of sound within it. This does not affect the
background evolution but is fundamental in determining a dark energyÕs
clustering properties, through the JeansÕ scale. It will, therefore, have an effect on
the evolution of fluctuations in the matter distribution.   

Following the papers laying the foundations for cosmological
perturbation theory \cite{Bardeen80, Kodama84}, the effect of the
speed of sound on observables was considered in more detail: for the
CMB  and large scale structure \cite{Hu98,Hu99} and in the context of
lensing \cite{Hu02} .  Observational implications of the speed of
sound in a variety of dark energy models have also recently been
discussed: for example for k-essence \cite{Erikson01,Dedeo03}, condensation of dark matter \cite{Bassett02} and the
Chaplygin gas, in terms of the matter power spectrum
\cite{Sandvik02,Beca03,Reiss03} and combined full CMB and large scale
structure measurements \cite{Bean03,Amendola03}.   

Minimally coupled scalar field, quintessence models, commonly have a
non-adiabatic speed of sound close or equal to unity (in units of c,
the speed of light),(see for example  \cite{Ferreira97}). By contrast
however, the adiabatic Chaplygin gas model,( \eg motivated by a
rolling tachyon \cite{Gibbons02}) has a speed of sound directly
proportional to the equation of state, both of which are approximately
zero up until late times when the dark energy starts to dominate. 
It is conceivable therefore that distinctions between such models
might be able to be made through the detection of a signature of the
dark energy speed of sound: in the large scale ISW correlations, and
in the cross correlation of the CMB with the distribution of large
scale structure \cite{Crittenden96}.   

In section \ref{sec2} we briefly review parameters describing a
general fluid and the issues that arise when establishing a fluid's
speed of sound. In section \ref{sec3} we describe the implications of
equation of state and speed of sound on perturbation evolution in the fluid and CDM. We
consider a toy model with slowly varying equation of state and sound
speed applicable in a wide variety of minimally coupled scalar field
theories. In section \ref{sec4} we discuss the potential for detecting the speed
of sound using late time perturbation evolution, in the ISW effect,
through the autocorrelation of the CMB temperature power spectrum. In
section \ref{sec5} we extend the analysis to the cross correlation of
the WMAP CMB data with distribution. Finally, in section \ref{sec6} we
summarize our findings. 

%%%%%%%%%%%%%%%%%%%%%%%%%%%%%%%%%%%%%%%%%%%%%%%%%%%%
\section{The speed of sound within general matter}\label{sec2}
%%%%%%%%%%%%%%%%%%%%%%%%%%%%%%%%%%%%%%%%%%%%%%%%%%%%

For a perfect fluid the speed of sound purely arises from adiabatic
perturbations in the pressure, $p$,  and energy density $\rho$ and the
adiabatic speed of sound, $c_{a}^{2}$ is purely determined by the
equation of state $w$, 
\bea
w_{i}&\equiv&{p_{i}\over\rho_{i}} \label{weq}\\
c_{ai}^{2}&\equiv&{\dot{p}_{i}\over \dot{\rho}_{i}}=w_{i}-{\dot{w}_{i}\over 3{\cal H}(1+w_{i})} \label{adisoseq}
\eea
where the subscript $i$ denotes a general specie of matter, where dots
represent derivatives with respect to conformal time and where ${\cal
  H}$ is the Hubble constant with respect to conformal time. 

In imperfect fluids, for example most scalar field or quintessence
models, however, dissipative processes generate entropic perturbations
in the fluid and this simple relation between background and the speed
of sound breaks down and we have the more general relation 
\bea
c_{si}^{2}&\equiv&{\delta p_{i}\over \delta\rho_{i}}.
\eea In order to establish the speed of sound in these cases we must look to the full action for the fluid 
described often through the form of an effective potential. In this case, the speed of sound can be written in terms of the contribution of the adiabatic component and an additional entropy
 perturbation $\Gamma_{i}$ and the density fluctuation in the given 
 frame $\delta_{i}$ \cite{Kodama84}, 

\bea
w_{i}\Gamma_i & \equiv & (c_{si}^{2} - c_{ai}^2)\delta_i  \\
&=& {\dot{p}_{i}\over\rho_{i}}\left({\delta p_{i}\over \dot{p}_{i}}-{\delta \rho_{i}\over \dot{\rho}_{i}}\right)
\label{Gammaeq}
\eea

$\Gamma_{i}$ is the intrinsic entropy perturbation of the matter
component, representing the displacement between hypersurfaces of
uniform pressure and uniform energy density.  In this paper, we are
solely interested in probing the intrinsic entropy of the dark energy
component. It is worth noting that in a multi-fluid scenario, in
addition to the intrinsic entropy perturbations denoted by
$\Gamma_{i}$, further contributions to the total entropy perturbation
of the system can arise from the relative evolution of two or more
fluids with different {\it adiabatic} sound speeds, and through
non-minimal coupling (see for example \cite{Balakin03}). 

Whereas the adiabatic speed of sound, $c_{ai}$, and $\Gamma_i$ are
scale independent, gauge invariant quantities, $c_{si}$ can be
neither. As such the general speed of sound is gauge and scale
dependent and issues of preferred frame arise. Looking at equation
(\ref{Gammaeq}), since the fluid $i$ rest frame is the only frame in
which $\delta_i$ is a gauge invariant quantity, this is the only frame in
which a matter component's speed of sound is also gauge-invariant. 

A useful transformation \cite{Kodama84} relates the gauge-invariant, rest frame density perturbation,  $\hat{\delta}_{i}$, to the density and velocity perturbations in a random frame, $\delta_{i}$ and $\theta_{i}$, 
\bea
\hat{\delta}_{i}=\delta_{i}+3{\cal H}(1+w_{i}){\theta_{i}\over k^{2}}\label{Deltaeq}
\eea
where we assume that the component is minimally coupled to other
matter species  and henceforth dark energy rest frame quantities are
denoted using a circumflex  (\^{}) . 

Using equations (\ref{Gammaeq}) and (\ref{Deltaeq}) we can rewrite the pressure perturbation in a general frame, $\delta p_{i}$,  in terms of the rest frame speed of sound,
\bea 
\delta p_{i} &=&\hat{c}_{si}^{2}\delta\rho_{i}+3{\cal
  H}(1+w_{i})(\hat{c}_{si}^{2}-c_{ai}^{2})\rho_{i}{\theta_{i}\over
  k^{2}}\label{deltapeq}. 
\eea

%%%%%%%%%%%%%%%%%%%%%%%%%%%%%%%%%%%%%%%%%%
\section{The speed of sound and perturbation evolution}\label{sec3}
%%%%%%%%%%%%%%%%%%%%%%%%%%%%%%%%%%%%%%%%%%
\begin{figure}[t]
\begin{center}
\includegraphics[width=3in]{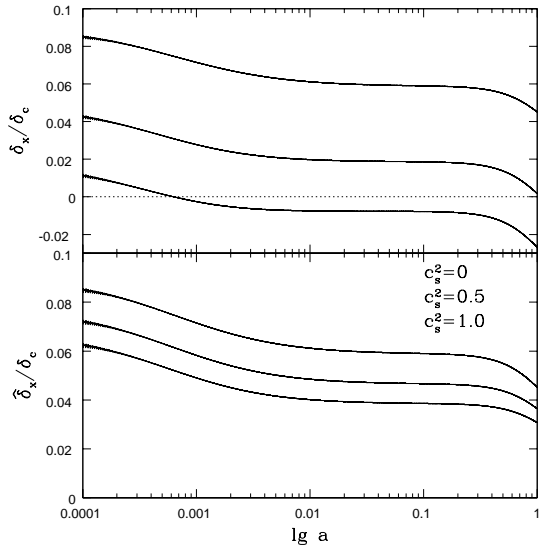}
\caption{Relative suppression of dark energy perturbations to those in CDM as one increases $\hat{c}_{s}^{2}$ from 0 to 1 (top to bottom in both panels) for $w=-0.8$. The top panel shows perturbations in CDM rest frame, which due to entropy perturbations can become negative. The bottom panel shows perturbations in dark energy rest frame which are always positive but for $w<0$ are a fraction of the CDM perturbations. $h$ and $\Omega_{x}$ are fixed so as to fit the WMAP constraints on $\Omega_{b}h^{2}, \Omega_{c}h^{2}$ and angular diameter distance to last scattering \cite{Spergel03}.} \label{fig1}
\end{center}
\end{figure}

Herein, we use the synchronous gauge and follow the notation of \cite{Ma95}. CDM rest frame quantities are denoted $\delta_{i}$ and $\theta_{i}$, while dark energy rest frame quantities use the circumflex (\^{}). The two are related by equation (\ref{Deltaeq}).

The energy density and velocity perturbation evolution of a general matter component in the CDM rest frame is given by
\bea \dot{\delta}&=&-(1+w)\left\{\left[k^{2}+9{\cal H}^{2}(\hat{c}_{s}^{2}-c_{a}^{2})\right]{\theta\over k^{2}}+{\dot{h}\over 2}\right\} \nonumber \\&& -3{\cal H}(\hat{c}_{s}^{2}-w)\delta\label{synceq3} 
\\
{\dot{\theta}\over k^{2}}&=&-{\cal H}(1-3\hat{c}_{s}^{2}){\theta\over k^{2}}+{\hat{c}_{s}^{2} \over 1+w}\delta\; .
\eea
 This set of equations illustrate clearly 
 that linear perturbations can be fully characterised by two numbers (and their potential time evolution): the equation of state and the rest frame speed of sound.
 
Let us now consider a toy model with a general fluid in which the time
variation in $w$ and $\hat{c}_{s}^{2}$ is small in comparison to the
expansion rate of the universe so that we can model it with constant
$w$  (i.e. $c_{a}^{2}\approx w$) and $\hat{c}_{s}^{2}$. Such models
are not impractical and can be used as the basis for comparison with
scalar field theories such as those with scaling potentials and
Chaplygin gases during the radiation and matter dominated eras. 

In the matter dominated era, ignoring baryons for simplicity, CDM density perturbations are affected by the speed of sound of dark energy (denoted `$x$') through the relation,
\bea
&&\ddot{\delta}_{c}+{\cal H}\dot{\delta}_{c}-{3{ \cal H}^{2}\over 2}\Omega_{c}\delta_{c}=  \\ && {3{ \cal H}^{2}\Omega_{x}\over 2}\left[(1+3\hat{c}_{s}^{2})\delta_{x}+9{\cal H}(1+w)(\hat{c}_{s}^{2}-w){\theta_{x}\over k^{2}}\right]\nonumber
\eea

In the radiation and matter dominated eras the expansion rate obeys an
effective power law $a\propto \tau^{m}$ and we find that the evolution
equations admit a solution of the form
$\delta_{c}\propto\delta_{x}/(1+w)\propto\tau^{2}$, and
$\theta_{x}\propto\tau^{3}$.

\begin{figure}[t]
\begin{center}
\includegraphics[width=3in]{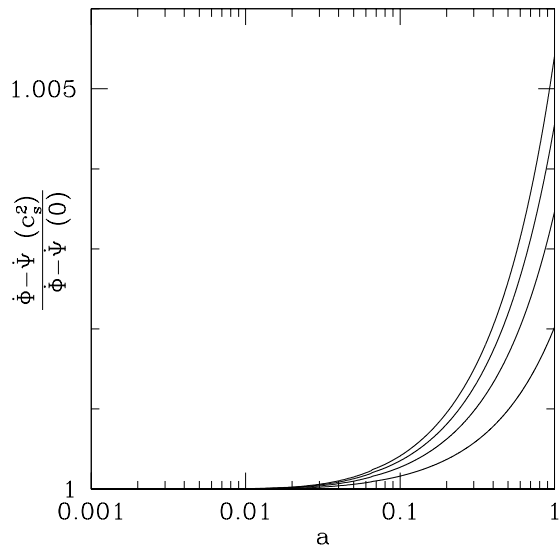}
\caption{Evolution the ISW source term, in comparison to the $\hat{c}_{s}^{2}=0$ scenario, with $w=-0.3$, for $\hat{c}_{s}^{2} = 0.25, 0.5, 0.75$ and 1 from bottom to top. $h$ and $\Omega_{x}$ are fixed as in Fig. \ref{fig1}. }
\label{fig2} 
\end{center}
\end{figure}

Equation (\ref{synceq3}) shows how $w$ and $\hat{c}_{s}^{2}$ affect the relative size of dark energy and CDM perturbations. In Fig. \ref{fig1} we see that, as one expects, increasing the speed of sound accentuates this suppression produced by reducing $w$, lowering $\delta_{x}/\delta_{c}$. Note that the dark energy density perturbation is well defined (remaining positive) in the rest frame, while the transformation into the CDM rest frame can make $\delta_{x}$ negative; this is just a foible of the frame one chooses, however.

The presence of dark energy perturbations leaves a $w$ and $c_{s}^{2}$ dependent signature in the ISW source term. This can be written in terms of the time variation of the anisotropic stress and the rest frame density perturbations of each matter component. 
\bea 
\dot{\Phi}-\dot{\Psi}
& \stackrel{z \rightarrow 0}{\approx}& -{1\over k^{2}}{d\over d\tau}\left[{\cal H}^{2}\left(\Omega_{c}\delta_{c}+\Omega_{x}\hat\delta_{x}\right)  \right]
\eea

Since $\hat{\delta}_{x}$ is suppressed in comparison to $\delta_{c}$, the dominant contribution to the ISW will come from the CDM perturbations. Subsequently it will be suppression of these (in comparison to a $\Omega_{c}=1$ scenario) through the effect of the dark energy speed of sound and equation of state that will leave a signature in the ISW. Fig 2. shows how as one increases $\hat{c}_{s}^{2}$ the ISW effect increases.

%%%%%%%%%%%%%%%%%%%%%%%%%%%%%%%%%%%%%%%%%%%%%%%%%%
\section{Constraints using WMAP temperature fluctuations at large scales}\label{sec4}
%%%%%%%%%%%%%%%%%%%%%%%%%%%%%%%%%%%%%%%%%%%%%%%%%%

\begin{figure}[t]
\begin{center}
\includegraphics[width=3in]{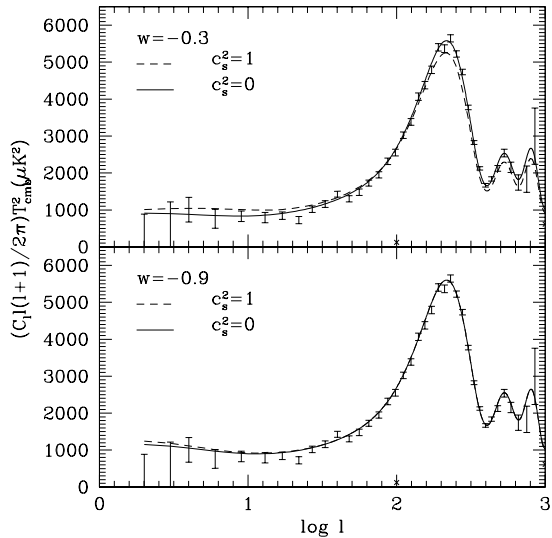}
\caption{CMB TT spectra for $w=-0.3$ (top panel) and $w=-0.9$ (bottom
  panel) with $\hat{c}_{s}^{2}=0$ and 1, all other parameters fixed to
  give the best fit at smaller scales. All spectra are normalized to
  $C_{87}$ for comparison.}  \label{fig3} 
\end{center}

\end{figure}
In this section, we investigate the joint constraints on the equation
of state and the speed of sound that can be infered using the CMB
temperature power spectrum. As was discussed earlier, the main effect
of a sound speed smaller than the speed of light will be felt at late
times and large scales and will thus only affect the very large scale
CMB temperature fluctuations arising from the late ISW effect, for
which WMAP already provide us will full sky cosmic variance limited
measurement \cite{Bennett03}. The Fourier component of the
fluctuations arising from the ISW effect are given by (for this
equation only, we ignore the well known issues related to the
sphericity of the observed sky). 
\bea
{\delta T\over T_{0}}(z,k)&=&\int^{\eta_{0}}_{\eta(z)}2\dot{\Phi}(\eta,k) d\eta 
\\
&=&{3H_{0}^{2}\Omega_{m}^{0}\over c^{2}k^{2}}\delta_{c}^{0}(k)\int^{z}_{0}{dg\over dz}dz
\eea
where $c^{2}$ is the square of the speed of light, $H_{0}$ is the
Hubble constant today, $\delta_{c}^{0}(k)=\delta_{c}(z=0,k)$ and
$\Omega_{m}^{0}$ is the fractional energy density in matter
(CDM+baryons) today and $g(z,k)=(1+z)D(z,k)$ where $D(z,k)$ is the
linear growth factor given by
$\delta_{c}(z,k)=D(z,k)\delta_{c}^{0}(k)$. In the limit that $w$ tends
to -1 $D(z,k)$ is scale independent and can be approximated by 
\bea
%D(z) & = & E(z)\left[1-{5\Omega_{c}^{0}\over 2\delta_{c}^{0}}\int_{0}^{z} {(1+z) dz \over E(z)^{3}} 
D(z) & = & {5\Omega_{c}^{0}E(z)\over 2}\int_{z}^{\infty} {(1+z) dz \over E(z)^{3}}\label{eq_D}\\
E(z) & = & {H(z)\over H_{0}}=\left[\sum_{i}\Omega_{i}^{0}(1+z)^{3(1+w_{i})}\right]^{1\over 2}.
\eea 
however this approximation does not provide the degree of precision
that is required for $w>-1$, even in the absence of dark energy
perturbations \cite{Linder03}. Because of this and in order to factor
in the late time effect of dark energy perturbations and their scale
dependence, for $\hat{c}_{s}^{2}\ne 0$, we explicitly calculate the
linear growth function, $D(z,k;\hat{c}_s^2,w)$ for each model. 

Note that the effect of the speed of sound comes in solely through the
value of $\delta_{c}^{0}$ while the equation of state affects both
$\delta_{c}^{0}$ and the linear growth factor. 
\begin{figure}[t]

\begin{center}
\includegraphics[width=3in]{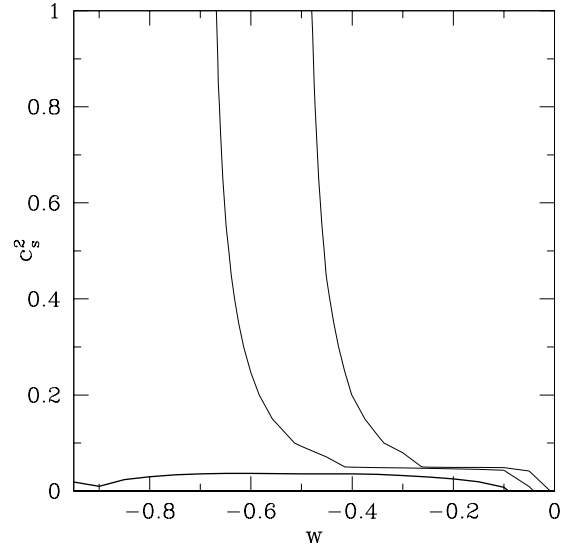}
\caption{Likelihood contour plot for the dark energy component in the $w-c_{s}^
{2}$ plane showing 1, 2 and 3$\sigma$ contours (heaviest to faintest
lines respectively) after marginalising over the power spectrum
normalisation .}  \label{fig4} 

\end{center}

\end{figure}

The associated auto power spectrum is given subsequently by
\bea
C_{\ell}^{TT}&=& 4\pi \int_{0}^{\infty} {dk\over k}\left\langle
{\delta T\over T_{0}}(k){ \delta T\over T_{0}}(k')\right\rangle
j_{\ell}^{2}(k\eta)\; . 
\eea

In order to probe solely the effect of $\hat{c}_{s}^{2}$ on
$C_{\ell}^{TT}$, we will compare to WMAP observations a family of
models lying along the angular diameter degeneracy surface present in
the CMB spectrum. To do so, we keep $\Omega_{c}h^{2}=0.135$ and
$\Omega_{b}h^{2}=0.0224$ to be consistent with the WMAP best fit
\cite{Spergel03} and choose $w$ and $h$ such that the angular diameter
distance to last scattering is the same. Other parameters correspond
to the best fit model of \cite{Spergel03} (table 7). Doing so, we
ensure that only the large scale correlations vary with each model and
that in all other respects they fit the WMAP data well. Note that we
also have vary slightly the overall amplitude due to the change in the
first peak height ISW plateau ratio. We consider $w$ values between 0
and -1 and $\hat{c}_{s}^{2}$ between 0 and 1. Given this grid of
model, we can then deduce easily the likelihood of the data using the
publically available code provided by the WMAP team
\cite{WMAPanalysis03}, from which we can deduce some joint constraints
on $w$ and $\hat{c}_s^2$. 

In Fig. \ref{fig3}  we show the variation of the CMB TT power spectrum
as one varies $\hat{c}_{s}^{2}$ from 0 to 1 for a model with $w=-0.3$
and -0.9. One can see that increasing $\hat{c}_{s}^{2}$ increases the
suppression of the CDM perturbations and therefore increases the power
on large scales. The effect decreases though as one decreases $w$; at
low $w$, the suppression due to the equation of state itself will
generate a dominant ISW effect on top of which a subdominant
contribution from $\hat{c}_{s}^{2}$ is then superimposed.  Those results
agree with the one obtained in \cite{Hu02}. 

In Fig. \ref{fig4} we show the likelihood plot from the WMAP data in
the $w-\hat{c}_{s}^{2}$ plane. The low quadrupole, and other low
$\ell$ $C_{\ell}$'s lead to a value of $\hat{c}_{s}^{2}<0.04$ being
preferred by the data, at the 1$\sigma$ level, although as one moves
to lower $w$ the ability to distinguish between different values of
the speed of sound disappears, because of cosmic variance.  

The cosmic variance thus limit our ability to constraint the dark
energy speed of sound using temperature $C_{\ell}$ only. However,
given the fact that all the constraints comes from the ISW effect,  it
is natural to consider the cross correlation the CMB with  the large
scale distribution of matter near us, correlation that is a direct
probe of the late ISW. In theory then, this might give us a better and
different probe into $\hat{c}_{s}^{2}$, so that both should be combine
eventually. We consider this in the next section. 

%%%%%%%%%%%%%%%%%%%%%%%%%%%%%%%%%%%%%%%%%%%%%%
\section{Constraints using CMB and large scale structure cross correlation}\label{sec5}
%%%%%%%%%%%%%%%%%%%%%%%%%%%%%%%%%%%%%%%%%%%%%%

\begin{figure}[t]

\begin{center}
\includegraphics[width=3in]{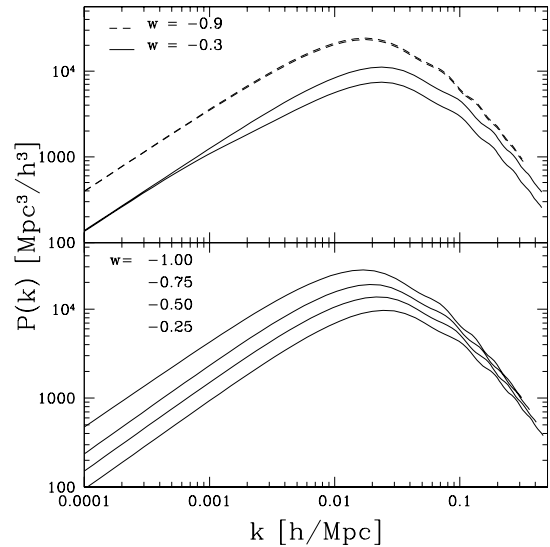}
\caption{Matter power spectra spectra COBE normalised\\
Top panel: $w=-0.9$ (dashed) and $-0.3$ (full) with $c_s^2=0$ and $1$ (top and bottom lines respectively )\\
Bottom panel:  $c_s^2=0$, $w=-1,-0.75,-0.5,-0.25$ from top to bottom (at large scales on both plots)}
\label{fig_Pk_cobe}

\end{center}

\end{figure}

As stated earlier, the dark energy affects very large scale modes of
dark matter density perturbations. As shown on Fig. \ref{fig_Pk_cobe}  
, those modes are outside the range of current wide field galaxy
surveys. For example, the SDSS measured the galaxy power spectrum down
to $\simeq$ 0.01 $h$Mpc$^{-1}$ "only" (see \eg
\cite{Dodelson01}). Full sky survey exists though, but their
particular properties and intrinsic limitations restricted their use
as direct probes of the matter power spectrum at those scales. For
example, the NRAO VLA Sky Survey (NVSS) \cite{Condon98} encompass such
a wide variety of objects that the difficulties in modeling the biases
at stake prevented its usage to directly measure (dark) matter density
fluctuations at those scales and infer this way any precise
cosmological constraint. However, their use in conjunction with large
scale CMB fluctuation measurements allows us to circumvent somehow
this difficulty. 

Indeed, within a given cosmological model, one can look at the
surveyed objects as a simple linearly bias tracer of dark matter
perturbations, a reasonnable approximation on those very large
scales. By measuring the auto-correlation function (ACF) of those
objects on those scales, one can infer the model dependant effective
bias for this composite population. Since this population trace the
large scale gravitational potential, it should correlate with the CMB
fluctuations induced by the same potential through the ISW effect. The
angular dependance of this cross-correlation function (CCF) and its
amplitude both depend on the tracer properties (bias and redshift
distribution) and on the particular cosmological model considered. In
particular, we would expect an important dependacy on the dark energy
properties which drive the evolution of the universe at those late
times and large scales. Modeling the tracer properties,  we can thus
in principle constrain the cosmology. This has been advocated first in
\cite{Crittenden96}, studied in details in \cite{Peiris00,Cooray02}
and performed effectively using as a tracer the NVSS sources
\cite{Boughn02b,Boughn03,Nolta03},  HEAO-1X-ray sources
\cite{Boughn98,Boughn02} or APM galaxies \cite{Fosalba03}. 

So far, this correlation has been probed to prove the very existence
of dark energy and to constrain its overall density. We here extend
this approach and try to investigate the potential constrains on its
very perturbative properties, \ie  jointlly its equation of state and
its sound speed. We will use as a data-set the ACF and CCF
measurements of \cite{Nolta03} performed using the NVSS catalog and
the WMAP 1-year maps. For the sake of simplicity we will follow the
same notations that we recall brievly. 

The measurements of fluctuations in the nearby matter distribution
from measuring the radio sources distribution can be expressed in
terms of the fractional source count perturbation given by, 
\bea
{\delta N\over
  N_{0}}(\hat{n})&=&b_{r}\delta_{c}^{0}(\hat{n})\int_{0}^{z}
{d\tilde{N}\over dz}D(z,{\hat n})dz
\eea
where $b_{r}$ is the linear bias in the matter distribution, $N_{0}$
is the mean source count per pixel (147.9 for 1.8$\deg$ square pixels
used in \cite{Nolta03}), and $d\tilde{N}/ dz$ is the normalised
redshift distribution of galaxies, such that $\int (d\tilde{N}/ dz)
dz=1$. For the latter we adopt the model of \cite{Dunlop90}. 

\begin{figure}[t]

\begin{center}
\includegraphics[width=3in]{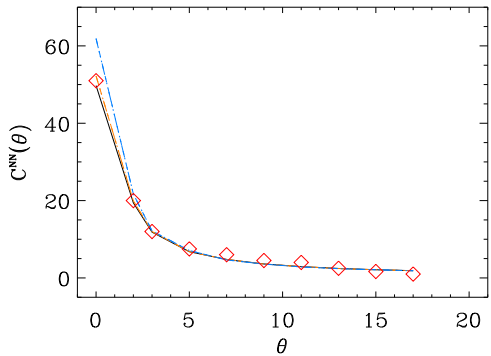}
\caption{Predicted ACF for the NVSS radio source catalog as a function
  of angular separation, $\theta$,, in degree. The linear bias factor
  has been accounted for. The various models correspond, in orange, to
  $w=-0.9$ and $c_s^2 = 0.9,0.1$ (respectively in triple dot-dashed or
  dot-dashed lines), or, in blue, to $w=-0.3$  with $c_s^2 = 0.9,0.1$
  (respectively in triple dot-dashed or dot-dashed lines). The black
  line corresponds to our fiducial $\Lambda$CDM model. Measurements in
  red are from \cite{Nolta03}.}  \label{fig_acf} 

\end{center}

\end{figure}

The dimensionless two point correlation function between two
quantities $X$ and $Y$ with background values $X_{0}$ and $Y_{0}$ in
positions   $\hat n$ and $\hat n'$ in the sky is given by 
\bea
C^{XY}(\theta)&=&\left\langle {\delta X\over X_{0}}(\hat n){ \delta Y\over Y_{0}}(\hat n')\right\rangle_{\hat n\cdot \hat n'=cos \theta} \nonumber \\&=& {1\over 4\pi}\sum_{\ell=2}^{\infty} (2l+1) C^{XY}_{\ell}P_{\ell}(cos \theta) \eea
For the fractional source count and CMB temperature cross and auto  
correlations  
\bea
C^{NT}_{\ell}&=& 4\pi \int_{0}^{\infty} {dk\over k}\left\langle {\delta N\over N_{0}}(k){ \delta T\over T_{0}}(k')\right \rangle j_{\ell}^{2}(k\eta) \nonumber\\
&=& 4\pi \int_{0}^{\infty} {dk\over k}\Delta^{2}(k)f_{\ell}^{N}(k)f_{\ell}^{T}(k)\; ,\\
C^{NN}_{\ell}&=& 4\pi \int_{0}^{\infty} {dk\over k}\left\langle {\delta N\over N_{0}}(k){ \delta N\over 
N_{0}}(k')\right \rangle j_{\ell}^{2}(k\eta) \nonumber\\
&=& 4\pi \int_{0}^{\infty} {dk\over k}\Delta^{2}(k)f_{\ell}^{N}(k)f_{\ell}^{N}(k)\; ,
\eea 
where 
\bea
\Delta^{2}(k)=\langle \delta_{c}^{0}(k)\delta_{c}^{0}(k')\rangle=\delta(k-k')k^{3}P(k)/2\pi^{2}
\eea
and where the filter functions for the temperature and number count fluctuations respectively are given by
\bea
f_{\ell}^{T}(k)&=& {3H_{0}^{2}\Omega_{m}^{0}\over c^{2}k^{2}}\int^{z}_{0}{dg(z,k)\over dz}j_{\ell}(k\eta(z))dz\\
f_{\ell}^{N}(k)&=&b_{r}\int_{0}^{z} {d\tilde{N}\over dz}D(z,k)j_{\ell}(k\eta(z))dz\; ,
\eea
where $D(z,k)$ has been previously defined in section \ref{sec4}.

\begin{figure}[t]

\begin{center}
\includegraphics[width=3in]{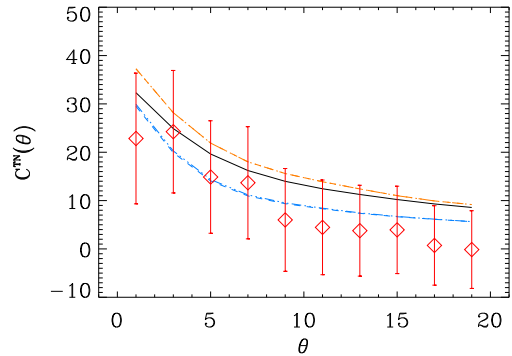}
\caption{Predicted CCF for the NVSS radio source catalog as a function
  of angular separation, $\theta$, in degree. The linear bias factor
  has been accounted for. The various models correspond, in orange, to
  $w=-0.9$ and $c_s^2 = 0.9,0.1$ (respectively in triple dot-dashed,
  dot-dashed lines), or, in blue, to $w=-0.3$  with $c_s^2 = 0.9,0.1$
  (respectively in triple dot-dashed or dot-dashed lines). They
  largely overlap. The black line corresponds to our fiducial
  $\Lambda$CDM model. Data points from \cite{Nolta03} are in red as
  well as (correlated) error bars.} 
\label{fig_ccf} 

\end{center}

\end{figure}
The results of this approach are summarised in figures \ref{fig_acf}
and \ref{fig_ccf}. As was used in section \ref{sec4}, we consider here
a family of models lying along the CMB angular diameter degeneracy
surface, and move along it by varying simultaneoulsy $w$ and $h$. For
each of these background, we then consider various $c_s^2$ and compute
the linear transfer function using a modified version of the CMBfast
\cite{cmbfast} or CAMB  \cite{camb} softwares . For a sample of those
models, we plotted both the predicted ACF, $C^{NN}(\theta)$,  from
which we infer the bias (figure \ref{fig_acf}) and the subsequent
predicted CCF, $C^{NT}(\theta)$ which one can compare with the measurements
of \cite{Nolta03} (figure \ref{fig_ccf}). Obviously, even if this
signal does indeed have some dependance with regards to the dark
energy pertubations parameters, $w$ and $c_s^2$, most of the effect is
absorbed in the bias measurement illustrated in figure
\ref{fig_acf}. The uncertainty in the bias is the main hindrance in
placing constraints on both the equation of state and speed of sound, 
so that current data do not allow this correlation is 
not measured well enough to constrain our models. We obtain a bias
range and a bias evolution similar as the one obtained by
\cite{Nolta03}, \ie $b_r \simeq 1.4$ for our fiducial $\Lambda$CDM
model, and $b_r$ tends to decrease when $w$ increases (in range
comprised between 1.4 and 2.2). Note that the plotted error bars are
heavily correlated. The knowledge of the correlation matrix computed
in \cite{Nolta03} allows us however to compute a well defined $\chi^2$
statistic. Note also that those error bars include cosmic variance
only but do not take into account the shot noise in the radio source
catalog.  It would be important to consider the uncertainties in the
bias measurement that propagates directly in the signal prediction but
we leave this issue for future work. Nevertheless, given the fact that
all our models lie within the one sigma error bars, we will not illustrate here by a
contour plot those weak joint constraints on $w$ and $c_s^2$.  

A natural and important question that arises at this level concerns
the future prospect for the measurements of this correlation,
depending on the used LSS tracers as well as the survey
considered. Although some studies have already been done
\cite{Hu99,Peiris00,Cooray02}, more specific investigations are
necessary. In particular an independent measurement of the bias, along
with improved scale and depth of survey will all contribute to vastly
improving on the current observational uncertainties. 

%%%%%%%%%%%%%%%%%%%%%%%%%%%%%%%%%%%%%%%%%%
\section{Conclusions}\label{sec6}
%%%%%%%%%%%%%%%%%%%%%%%%%%%%%%%%%%%%%%%%%%

We have reviewed the effect of the speed of sound of dark energy on
CDM and dark energy perturbations. While a positive dark energy speed
of sound suppresses the CDM perturbations, it is the deviation from
adiabaticity, in combination with the equation of state that determine
the degree of suppression of the amplitude dark energy perturbation in
comparison to those of the CDM. 
 
We have found the CMB large scale temperature fluctuations, dominated
by the ISW effect, are a promising tool to measure the speed of
sound. The suppression of CDM matter perturbations drive the late time
ISW effect.  

From the auto correlation   of the WMAP data with itself we obtain a
$1\sigma$ "constraint" on the speed of sound $\hat{c}_{s}^{2}<0.04$,
using scenarios that minimise contributions to the likelihood on small
scales (from the peaks) as much as possible by using well known
degeneracies to follow the WMAP best fit model as closely as possible.
The main limitation in obtaining constraints from the auto correlation
is the cosmic variance.  

We have also investigated the cross correlation of the large scale CMB
with fluctuations in the nearby mass distribution using the NVSS radio
source catalogue. We here again find that cosmic variance is a strong
limitation and prevent us from placing any strong constraint in the
$w-c_s^2$ plane. 

However, since the potential of such an analysis might be unique in
unveiling the mysteries of the dark energy,  it is important to
explore further out the prospect of future potential large scale probe
of the gravitational portential and so of the ISW (LSST, PLANCK,
CMBPOL). We have presented some estimates of prospective
  constraints that one might obtain from cross correlation of large
  scale probes with CMB however we leave this exploration for 
  future work. 

%%%%%%%%%%%%%%%%%%%%%%%%%%%%%%%%%%%%%%%%%%
\medskip
{\bf Acknowledgements} We would like to thank Robert Caldwell, Sean
Carroll, Anthony Challinor, Rob Crittenden, Joe Hennawi, Mike Nolta, Hiranya
Peiris, Martin White and, especially, David Spergel for very helpful discussions and questions in the course of this work. O.D. acknowledges the Aspen Center for Physics where  
part of this work was pursued. R.B. and O.D. are supported by WMAP and
NASA ATP grant NAG5-7154 respectively.  
%%%%%%%%%%%%%%%%%%%%%%%%%%%%%%%%%%%%%%%%%%


\begin{thebibliography}{99}

\bibitem{Bennett03} C. L. Bennett \etal astro-ph/0302207 
\bibitem{super} P.M. Garnavich et al, Ap.J. Letters \textbf{493}, L53-57 (1998); S. Perlmutter
et al, Ap. J. \textbf{483}, 565 (1997); S.  Perlmutter et al (The SupernovaCosmology Project), Nature \textbf{391} 51 (1998); A.G. Riess et al,Ap. J. \textbf{116}, 1009 (1998)
\bibitem{Spergel03} D. N. Spergel \etal astro-ph/0302209
\bibitem{Efstathiou03} G.~Efstathiou, astro-ph/0303127.
\bibitem{Contaldi03} C.R. Contaldi, M. Peloso, L. Kofman, A. Linde, astro-ph/0303636.
\bibitem{Efstathiou03b} G.~Efstathiou, astro-ph/0306431
\bibitem{Kofman85} L.~A. Kofman \& A.~A. Starobinskii\ 1985, Soviet Astronomy Letters, 11, 271 
\bibitem{Crittenden96} R.G. Crittenden, N. Turok \PRL {\bf 76} 4 (1996).
\bibitem{Boughn03} S.P. Boughn, R.G Crittenden astro-ph/0305001
\bibitem{Nolta03} M.R. Nolta \etal astro-ph/0305097
\bibitem{Fosalba03} P. Fosalba \& E. Gatzanaga, astro-ph/0305468

\bibitem{Peebles02} P. J. E. Peebles, B. Ratra, RMP (2003) in
press, astro-ph/0207347.

\bibitem{Bardeen80} J.M. Bardeen, \PR D {\bf 22} 1882 (1980)
\bibitem{Kodama84} H. Kodama \& M. Sasaki, {\PTP} {\it Supp.} {\bf 78} 1 (1984)
\bibitem{Hu98} W. Hu, \ApJ {\bf 506} 485H (1998), astro-ph/9801234
\bibitem{Hu99} W. Hu., D.~J. Eisenstein, M. Tegmark \& M. White \ 1999, \prd, 59, 23512 
\bibitem{Hu02} W. Hu,\PR D {\bf 65} 023003 (2002), astro-ph/0108090.
\bibitem{Erikson01} J. Erikson, R.R. Caldwell, P.J. Steinhardt, V. Mukhanov, and C. Armendariz-Picon \PRL {\bf 88} 121301 (2001)
\bibitem{Dedeo03} S. DeDeo, R.R. Caldwell, P.J. Steinhardt, astro-ph/0301284
\bibitem{Bassett02} B. A. Bassett, M. Kunz, D. Parkinson, C. Ungarelli, astro-ph/0210640; astro-ph/0211303
\bibitem{Sandvik02} H. Sandvik, M. Tegmark, M. Zaldarriaga, I. Waga,
astro-ph/0212144.
\bibitem{Beca03} L.M.G. Beca, P.P. Avelino, J.P.M. de Carvalho, C.J.A.P. Martins
astro-ph/0303564, to appear in \PR D
\bibitem{Reiss03} R.R.R Reiss, I. Waga, M.O. Calv\~{a}o, S.E. Jor\'{a}s astro-ph/0306004.
\bibitem{Bean03} R. Bean, O. Dor\'{e}, \PR D in press, astro-ph/0301308
\bibitem{Amendola03}L. Amendola, F. Finelli, C. Burigana, D. Carturan astro-ph/0304325
\bibitem{Ferreira97} P.G. Ferreira, M. Joyce, \PRL {\bf 79} 4740 (1997); P.G. Ferreira, M. Joyce, \PR D {\bf 58}  023503 (1998)
\bibitem{Gibbons02} G. Gibbons, \PL  B {\bf 537} 1 (2002), hep-th/0204008
\bibitem{Balakin03} A. B. Balakin, D. Pav\'{o}n, D. Schwarz, W. Zimdahl, astro-ph/0302150

%\bibitem{Boughn02} S.P. Boughn, R.G Crittenden \PRL {\bf 88} 2 (2002).

\bibitem{Ma95} C.P Ma, E. Bertschinger \ApJ {\bf 455} 7 (1995).
\bibitem{Linder03} E.V. Linder, A. Jenkins astro-ph/0305286
\bibitem{WMAPanalysis03} Verde et al., astro-ph/0302218,  Hinshaw et al. astro-ph/0302217; A.~Kogut {\it et al.}, astro-ph/0302213.
\bibitem{Dodelson01} S. Dodelson \etal,  \ApJ, 572, 140-156 (2001)
\bibitem{Condon98} J. Condon \etal, {\it Astron. J.} 115, 1693 (1998)
\bibitem{Peiris00} H. Peiris \& D. Spergel, \apj, 540, 605, 2000
\bibitem{Cooray02} A.  Cooray, A.\ 2002, \prd, 65, 103510
\bibitem{Boughn02b} S.~Boughn, \&  R.~Crittenden, \PRL, 88, 21302 (2002)
\bibitem{Boughn98} S.~Boughn, R.~Crittenden, \& N.~Turok\ 1998, New Astronomy, 3, 275 
\bibitem{Boughn02} S. Boughn, R.~Crittenden, \& G.~P. Koehrsen, \apj, 580, 672 (2002)
\bibitem{Dunlop90} J.S. Dunlop, J.A. Peacock {\it MNRAS} {\bf 247} 19 (1990).
\bibitem{cmbfast} Seljak, U.~\& Zaldarriaga, M.\ 1996, \apj, 469, 437 , see also \texttt{http://cmbfast.org/}
\bibitem{camb} \texttt{http://camb.info/}





\end{thebibliography}
\end{document}